\documentclass[10pt]{iopart}
\usepackage{ulem}
\usepackage{textcomp}
\usepackage{graphicx}
\usepackage{subfig}
\usepackage[usenames,dvipsnames]{xcolor}
\usepackage{url}
\usepackage{citesort}
\usepackage{color,soul}

\providecommand{\jt}[1]{\textcolor{black}{#1}}

\providecommand{\ue}[1]{\textcolor{black}{#1}}
\begin{document}
\title[]{Simulation of positive streamers in CO$_2$ and in air:  the role of photoionization or other electron sources}
\author{Behnaz Bagheri$^{1,2}$, Jannis Teunissen$^{1}$, and Ute Ebert$^{1,2}$}
\address{$^1$Centrum Wiskunde \& Informatica (CWI), Amsterdam, The Netherlands
\\$^2$Technical University of Eindhoven, Eindhoven, The Netherlands}
\ead{b.bagheri@tue.nl}
\vspace{10pt}
\begin{indented}
\item[]\today
\end{indented}

\begin{abstract}
Positive streamer discharges have been studied and modelled extensively in air.
Here we study positive streamers in CO$_2$ with and without oxygen admixtures; they are relevant for current high voltage technology as well as for discharges in the atmosphere of Venus. We discuss that no efficient photoionization mechanism is known for gases with a large CO$_2$ fraction, as photons in the relevant energy range are rapidly absorbed. Hence positive streamers can propagate only due to some other source of free electrons ahead of the ionization front. Therefore we study positive streamer propagation in CO$_2$
  % and in CO$_2$ with admixtures of O$_2$
  with different levels of background ionization to provide these free electrons. The effect of replacing photoionization by background ionization is studied with simulations in air.
Simulating streamers in background fields of 16 to 20~kV/cm at standard temperature and pressure within a gap of 6.4~cm, we find that streamer propagation is rather insensitive to the level of photoionization or background ionization. 
We also discuss that the results depend not only on the value of breakdown field and applied electric field, and on preionization or photoionization, but also on the electron mobility $\mu(E)$ and the effective ionization coefficient $\alpha_{\rm eff}(E)$, that are gas-dependent functions of the electron energy or the electric field.
\end{abstract}
%\linespread{1.5}
%\onecolumn

\ioptwocol
\section{Introduction}
\label{sec:introduction}

\subsection{Positive streamers in air and other N$_2$:O$_2$ mixtures}

%\sout{HERE MORE REFERENCES TO PREVIOUS WORK CAN BE INSERTED.}\\
Streamers are rapidly growing ionized filaments which govern the initial phase of electric breakdown; they later can develop into a spark or a lightning leader~\cite{vitello_simulation_1994,yi_experimental_2002,ebert_review_2010,Nijdam_review_2020}. Their growth is governed by the curved space charge layer around their tips which enhances the electric field in the non-ionized areas in front of them and allows them to penetrate into areas where the background electric field is below the breakdown threshold. They are weakly ionized channels, hence they do not increase the gas temperature significantly. We focus here on positive streamers that \ue{start and propagate more easily in air than negative streamers}. Positive streamers propagate in the direction of the electric field with velocities comparable to the local electron drift velocity, but in the opposite direction, therefore they require a source of free electrons in front of their head to sustain their growth. In N$_2$:O$_2$ mixtures like air, these electrons are provided by photoionization~\cite{nijdam_probing_2010,pancheshnyi_role_2005}. Background ionization, e.g., from previous discharges, can further influence their growth. 
\ue{In the present paper, the role of photoionization or other electron sources 
for positive streamer propagation will be investigated, in particular, in CO$_2$ with or without admixtures of other gases.}

Streamers are used in numerous applications in plasma technology~\cite{Fridman_2005,Adamovich_2017}, for instance for the production of
chemical radicals~\cite{kanazawa_observation_2011}, in ignition and
combustion~\cite{starikovskaia_plasma-assisted_2014} and in plasma catalysis
\cite{nozaki_non-thermal_2013}. 
%~\cite{Cernak_2011,starikovskaia_2014,Starikovskii_2009,Joshi_2013,jiang_2014,nozaki_2013,Akiyama_2000,Kanazawa_2011,Nastuta_2011,Sands_2008,Bekeschus-2016-ID4126}.

\subsection{Positive streamers in CO$_2$ with or without admixtures}

In the present work, we concentrate on properties of positive streamers in CO$_2$. The study is motivated by current needs in high-voltage technology, where pressurized gas is used for insulation and current interruption purposes~\cite{nakanishi1991switching,ryan1989sf6,Seeger_2016}.
The commonly used working gas in high voltage circuit breakers and many other applications in high voltage technology is Sulphur Hexafluoride~(SF$_6$) due to its unique insulating properties. However, it is a strong green house gas with a global warming potential of 23900 times that of CO$_2$ on a 100 year horizon~\cite{UN-2014}. Furthermore, it produces highly toxic components under switching operation. 

The search for an alternative gas has revealed that CO$_2$ is a suitable replacement for SF$_6$~\cite{Seeger_2016,singhasathein2013dielectric,okubo2011recent,uchii2004investigations,uchii2004development,uchii2004fundamental,seeger2015perspectives,stoller2013rm}. ABB has introduced the first high voltage circuit breaker using CO$_2$ to the market~\cite{ABB-switchgear}. The typical pressure range of CO$_2$ is then around~1~-~10 bar~\cite{Seeger_2016}. 

Knowledge on discharge dynamics in CO$_2$ is also relevant for lightning on Venus that has an atmosphere of 96.5\% CO$_2$ and 3.5\% N$_2$. While no optical signature of lightning activity has been reported (probably due to the low luminosity of CO$_2$ discharges in the visible range, or due to the opacity of the dense Venus atmosphere), electromagnetic remote sensing indicates lightning at a similar frequency as on earth~\cite{dubrovin2010sprite}.  

Experimental investigations of streamer stability field, streamer radius and streamer velocities in CO$_2$ at ambient temperature in the pressure range of $0.5-5$~bar for both positive and negative polarities are presented in~\cite{Seeger_2016}. On the other hand, %to the best of our knowledge, 
simulations of positive streamers in CO$_2$ have only been performed in~\cite{levko_particle--cell_2017};
the authors used 2D Cartesian particle-in-cell Monte Carlo and 2D Cartesian fluid simulations to study streamer branching; however a 2D Cartesian computation gives only a qualitative picture. In contrast, here we study the propagation of streamers in CO$_2$ with axisymmetric fluid simulations, and we study how the streamer properties depend on the gas composition, electric field and background electron density.

%\sout{SUGGESTED REPLACEMENT PARAGRAPH:}\\
A major bottleneck in the study of positive streamers in pure CO$_2$ is that there seems to be no effective photo-ionization in that gas, as we review and discuss in section~\ref{sec:photoi-model} of this paper.
This is true as well for CO$_2$ \ue{with admixtures of other gases, in particular,} of O$_2$ or of air. Therefore streamer propagation in CO$_2$ seems only understandable with some other source of free electrons ahead of the streamer,
e.g., due to radiation or to previous discharges. 
Therefore we insert different values of background electron densities and study their effect on the streamer propagation.
\ue{Without such an electron source, streamer inception in CO$_2$ will be difficult and streamer propagation erratic with multiple branching attempts, as is discussed further in section~\ref{sec:posit-stre-co2}.}

\subsection{Positive streamers in other gases}

Positive streamers depend essentially on three functions that are specific for the particular gas composition:
%In particular, three factors affect streamer properties: 
on the electron mobility $\mu(E)$, on the effective ionization coefficient $\alpha_{{\rm eff}}(E)$, and on the distribution of photoionization or possibly some other source of free electrons ahead of the ionization front. The breakdown field is defined as the field where $\alpha_{{\rm eff}}(E)=0$. But beyond this single value that sets a scale for the electric field, the functional dependence of electron mobility $\mu$ and effective ionization coefficient $\alpha_{\rm eff}$ on the electric field $E$ \ue{or the electron energy} determines streamer properties like \ue{velocity, radius and maximal electric field at the tip, and electric field and electron density in the streamer interior.}
To compare how streamer properties depend on these functions, in this paper, we study streamers in air, in CO$_2$, and in CO$_2$ with $1\%$ or $10\%$ admixture of O$_2$ at standard temperature and pressure. (Note that other gas densities with the same mixture ratios can be approximated by scaling laws~\cite{ebert_review_2010,briels_positive_2008-1}).

\subsection{Contents of the paper}
The structure of paper is as follows: 
Section~\ref{sec:discharge-model} is devoted to the plasma fluid model with initial and boundary conditions. Subsection~\ref{sec:photoionization} reviews the literature on photoionization in air, in pure CO$_2$, and in CO$_2$ with admixtures of oxygen, air and other gases. Transport and reaction parameters for air and for CO$_2$ with or without admixture of oxygen are provided in section~\ref{sec:transport-reaction}. In section~\ref{sec:results}, first the difficult propagation of CO$_2$ streamers without background ionization is discussed, and a first view on streamers in air and in CO$_2$ is given. Then the effect of replacing photoionization by background ionization in air streamers is discussed in section~\ref{sec:air}. In section~\ref{sec:CO2}, we characterize CO$_2$ streamers, and in section~\ref{sec:airvsCO2} streamers in air are compared to those in CO$_2$. The effect of oxygen admixture of $1\%$ or $10\%$ on CO$_2$ streamers are presented in section~\ref{sec:CO2-O2}. Finally, the concluding remarks are presented in section~\ref{sec:conclusions}.

\section{Discharge model and conditions}
\label{sec:discharge-model}
We use a plasma fluid model for the densities of electrons and ions that incorporates elastic and inelastic collisions of electrons with O$_2$, N$_2$ and CO$_2$ molecules, including impact ionization and electron attachment reactions. Details on the calculation of mobility and reaction rates are given in section~\ref{sec:transport-reaction}.
%which contains electrons and the following ions: N$_2^{+}$, CO$_2^{+}$, O$_2^{+}$, O$_2^{-}$ and O$^{-}$. 
%We also trace the density of excited nitrogen molecules N$_2^{*}$ relevant for photo-ionization. The drift-diffusion approximation is used for electrons, whereas
% Jannis removed: The electrons drift and diffuse in a self-consistent electric field.
Ions are considered immobile during the initial streamer phase due to their larger mass. For each gas we consider the respective reactions that are listed in table~\ref{tab:reaction_table}. 
%The source of photons for the photo-ionization process is also included as a reaction in our model. 
Note that the rates of the three-body attachment reaction, e + O$_2$ + M $\rightarrow$ O$_2^{-}$ + M with M~=~N$_2$, are about three orders of magnitude smaller than the respective rates for  M~=~O$_2$~\cite{kossyi_kinetic_1992}. Therefore, we only consider the three-body attachment with M~=~O$_2$.

The model is implemented in Afivo-streamer~\cite{teunissen_simulating_2017}. It is based on the Afivo
framework~\cite{Teunissen_afivo_2018}, which contains geometric multigrid
techniques to solve the Poisson equation, octree-based adaptive mesh
refinement~(AMR) and OpenMP parallelism. The fluid equations are solved using explicit
second order time stepping, and a slope-limited second order accurate spatial
discretization.

\begin{table}
\begin{center}
\begin{tabular}{ c  c  c}
\hline
1 & e + CO$_2$ $\rightarrow$ e + e + CO$_2^{+}$ &  $k_1(E/N)$ \\ 
2 & e + N$_2$ $\rightarrow$ e + e + N$_2^{+}$ & $k_2(E/N)$ \\  
3 & e + O$_2$ $\rightarrow$ e + e + O$_2^{+}$ & k$_3(E/N)$    \\
 4 & e + O$_2$ + O$_2$ $\rightarrow$ O$_2^{-}$ + O$_2$ & k$_4(N,~E/N)$\\
 5 & e + O$_2$ $\rightarrow$ O$^{-}$ + O & k$_5(E/N)$\\
 6 & e + CO$_2$ $\rightarrow$ CO + O$^{-}$ & k$_6(E/N)$\\
\\
 \hline
\end{tabular}
\caption{List of ionization and attachment reactions in the model used for the different gas compositions. The reaction rates depend on the reduced electric field $E/N$ (where $E$ is the field and $N$ the gas density). They are calculated from elastic and inelastic cross sections as explained in section~\ref{sec:transport-reaction}, and they are provided as input files to the fluid model. Reaction~4 also depends on the O$_2$ density, or on the gas density~$N$, when assuming a constant fraction of O$_2$.}
\label{tab:reaction_table} 
\end{center}
\end{table}

\subsection{Model equations}
The electron density $n_e$ evolves in time as
\begin{equation}
  \label{eq:dt-ne}
 \partial_t n_e=\nabla \cdot (n_e\mu_e\mathbf{E} + D_e\nabla n_e) + S_i + S_\mathrm{ph} - S_{\rm {attach}},
\end{equation}
where $\mu_e$ is the (positive) electron mobility, $D_e$ the electron
diffusion coefficient, $\mathbf{E}$ the electric field, $S_i$ the impact ionization source term, $S_{\rm {attach}}$ the electron attachment source term, and $S_\mathrm{ph}$ the non-local photo-ionization source
term (see section \ref{sec:photoi-model}). All mobility and reaction coefficients are 
calculated in local field approximation.
%$\bar{\alpha}$ the effective ionization coefficient,
%$\mathbf{E}$ the electric field, $S_i = \bar{ \alpha}\mu_e |\mathbf{E}|n_e$ the
%ionization source term and $S_\mathrm{ph}$ the non-local photoionization source
%term (see section \ref{sec:photoi-model}).
%\begin{eqnarray}
%  \partial_t n_e &= \nabla \cdot (n_e\mu_e\mathbf{E} + D_e\nabla n_e)
%                   + S_i + S_\mathrm{ph} + K_\mathrm{e},\\
% % \partial_t n_i &= S_i + S_\mathrm{ph},
% S_i &= n_e [CO_2]k_1 + n_e [N_2]k_2+n_2[O_2]k_3\\
% K_e &= \left( n_e[O_2]^2k_4 + n_e[O_2]k_5 +n_2 [CO_2]k_6  \right)
%  \label{equ:fluid-model}
%\end{eqnarray}
\\
The total positive ion density~$n_i^{+}$ and the total negative ion density~$n_i^{-}$ change in time as
\begin{eqnarray}
\partial_t n_i^{+} &= S_i + S_{\rm{ph}},\\
\partial_t n_i^{-} &= S_{\rm {attach}}.
\end{eqnarray}
%Note that in this paper we do not need to distinguish between positive and negative ions. The ion density is only used to compute the space charge density. \hl{Jannis: I would remove this last remark.}

%DISCUSS THAT YOU DO NOT NEED TO DISTINGUISH THE DIFFERENT TYPES OR POSITIVE OR NEGATIVE IONS. YOU ACTUALLY COULD CALCULATE WITH JUST ONE ION DENSITY AS LONG AS YOU DON'T INCLUDE ANY FURTHER REACTIONS OF THE IONS. 

%\begin{eqnarray}
%\partial_t [\mathrm{N}_2^{+}] &= n_e [\mathrm{N}_2] k_2,\\
%\partial_t [\mathrm{CO}_2^{+}] &= n_e [\mathrm{CO}_2] k_1,\\
%\partial_t [\mathrm{O}_2^{+}] &= n_e [\mathrm{O}_2] k_3 + S_\mathrm{ph},\\
%\partial_t [\mathrm{O}_2^{-}] &= n_e [\mathrm{O}_2]^2 k_4 + n_e [\mathrm{O}_2] k_5,\\
%\partial_t [\mathrm{O}^{-}] &= \partial_t [\mathrm{CO}] = n_e [\mathrm{CO}_2]k_6,
%%\partial_t [\mathrm{O}^{-}] &= n_e [\mathrm{CO}_2]k_6,
%\end{eqnarray}
%in which [...] indicates the density of the respective species, $k_j, j=1,2,...,6$ the respective reaction rates.
The electric field is computed in electrostatic approximation as
\begin{eqnarray*}
  \mathbf{E} &= -\nabla \phi,\\
  \nabla^{2}\phi &= -\frac{q}{\epsilon_0},\quad q = {\rm e}\left(n_i^+-n_i^--n_e\right),
\end{eqnarray*}
where $\phi$ is the electric potential, $\epsilon_0$ the vacuum permitivity,
$q$ the space charge density and e the elementary charge.
The impact ionization and the electron attachment source terms are computed according to
\begin{eqnarray}
S_i &= n_e [{\rm CO}_2]k_1 + n_e [{\rm N}_2]k_2+n_e[{ \rm O}_2]k_3, \label{eq:Si}\\
 S_{\rm{attach}} &= n_e[{\rm O}_2]^2k_4 + n_e[{\rm O}_2]k_5 +n_e [{\rm CO}_2]k_6,
%\label{equ:fluid-model}
\end{eqnarray}
where [...] indicates the density of the respective species, and $k_j, j=1,2,...,6$ are the respective reaction rates that still depend on the specific gas composition,
as discussed further in section \ref{sec:transport-reaction}. For further reference, we recall that the ionization energies of O$_2$, N$_2$ and CO$_2$ are 12.1~eV, 15.6~eV and 13.8~eV~\cite{pancheshnyi_photoionization_2014}.

\subsection{Photoionization}
\label{sec:photoionization}
\subsubsection{Air}
\label{sec:photoi-model}
For positive streamers in air, photoionization provides the free electrons in front of the streamer head~\cite{nijdam_probing_2010,pancheshnyi_role_2005} that are needed for streamer propagation into gases without preionization. 
%The streamer path is influenced by the location of free electrons as it is propagating.
It is generally accepted that photoionization in air occurs when excited nitrogen molecules emit radiation, which is absorbed by oxygen molecules and ionizes them. According to Zheleznyak~{\it et al}~\cite{zheleznyak_photoionization_1982} the wavelength of such radiation is in the 98 to 102.5~nm range; in this band the photon energy exceeds the ionization energy of 12.1~eV of O$_2$, and the photon absorption by nitrogen molecules is negligible.

When the photons are emitted isotropically and not scattered in the medium, and when the photon travel time is negligible, the photo-ionization source term $S_{\mathrm{ph}}(\mathbf{r})$ can be written for each photon wave length as
\begin{equation}
S_{\mathrm{ph}}({\bf r})=\int d^3{\bf r}'\;\frac{I({\bf r'})f(|{\bf r}-{\bf r'}|)}{4\pi|{\bf r}-{\bf r'}|^2}.
\label{equ:photo-general}
\end{equation}
Here $I(\mathbf{r})$ is the source of ionizing photons,
$4\pi|\mathbf{r}-\mathbf{r}'|^2$ is a geometric factor, and $f(r)$ is
the absorption function. 

In Zheleznyak's model an effective function $f(r)$ for the wave length range of
98 to 102.5~nm is given by
 \begin{equation}
f(r)=\frac{\exp(-\chi_{\mathrm{min}}p_{O_2}r)-\exp({-\chi_{\mathrm{max}}}p_{O_2}r)}{r\ln(\chi_{\mathrm{max}}/\chi_{\mathrm{min}})},
\label{equ:absorption-function}
\end{equation}
where $\chi_{\mathrm{max}}\approx1.5\times10^2/(\textnormal{mm bar})$,
$\chi_{\mathrm{min}}\approx2.6/(\textnormal{mm bar})$, and $p_{O_2}$ is the
partial pressure of oxygen. Zheleznyak's UV photon source term
$I(\mathbf{r})$ is proportional to the electron impact ionization source term $S_i$
as given in equation (\ref{eq:Si})
\begin{equation}
I({\bf r})=\frac{p_q}{p+p_q}\xi S_i,
\label{equ:Zhelez-source-term}
\end{equation}

where the factor $p_q/(p+p_q)$ accounts for the collisional quenching of the excited
nitrogen molecules, where $p$ is the actual gas pressure 
and $p_q$ a gas specific quenching pressure. %$p_q = 40 \, \mathrm{mbar}$. 
In air at standard temperature and pressure, the corresponding absorption lengths are $[\chi_{\mathrm{min}}p_{O_2}]^{-1}=1.9$~mm and $[\chi_{\mathrm{max}}p_{O_2}]^{-1}=33~\mu$m.
The proportionality factor $\xi$, which relates the impact excitation to the impact ionization, is in
principle field-dependent \cite{zheleznyak_photoionization_1982}, but in this paper, we set it to $\xi = 0.05$. Furthermore, we use a quenching
pressure of $p_q = 40 \, \mathrm{mbar}$.

Having the UV photon source term calculated, we evaluate the integral in equation~(\ref{equ:photo-general}) by using a set of Helmholtz differential equations~\cite{bourdon_efficient_2007,luque_photoionization_2007} with Bourdon's three-term parameters~\cite{bourdon_efficient_2007}. Besides to the original papers, the reader is referred to \cite{bagheri2019effect} and the appendix of  \cite{bagheri_comparison_2018} for more details.

\subsubsection{CO$_2$}
\label{sec:photoi-model-CO2}
Even though there have been many studies on the physics of photoionization in air as well as on its numerical implementation in discharge models~\cite{pancheshnyi_photoionization_2014,luque_photoionization_2007,bourdon_efficient_2007,Stephens_2016,Stephens_2018}, to the best of our knowledge there are no quantitative photoionization models for discharges in CO$_2$ and CO$_2$ containing gas mixtures. 

Direct measurements of absorption coefficients in CO$_2$ were only reported by Przybylski~\cite{przybylski_1962} and Teich~\cite{teich_emission_1967} and retrieved by Pancheshnyi~\cite{pancheshnyi_photoionization_2014}; they are in the range of $0.34~-~2.2~\rm{cm}^{-1}\rm{Torr}^{-1}$~($=25-165$~$\rm{mm}^{-1}\rm{bar}^{-1}$),
which at standard temperature and pressure correspond to absorption lengths in the range of $6.1~-~40~\mu$m. Pancheshnyi~\cite{pancheshnyi_photoionization_2014} attributed these values to the spectral range of $83~-~89$~nm, emitted by
the dissociative ionization excitation products of CO$_2$~\cite{sroka_light_1970,van_der_burgt_photoemission_1989}. The lower energy threshold to generate such products is about 50~eV~\cite{sroka_light_1970}, hence not in a relevant energy regime for electrons in typical streamer discharges. Therefore, we expect negligible photoionization in pure CO$_2$. 

The authors of~\cite{levko_particle--cell_2017} also neglected photoionization of CO$_2$, but with a different argument than above; and they included a quasi neutral plasma with density of $10^{9}$~$/\rm{m}^3$ in their simulation.

\subsubsection{CO$_2$ with admixtures of oxygen or air}
\label{sec:photoi-model-CO2-air}
In commercial circuit breakers based on CO$_2$, there are  various contaminations with air and other impurities as well as admixtures of other gases. In particular, an admixture of O$_2$ is used in a breaker to suppress soot formation from the CO$_2$ discharge.

Photoionization of CO$_2$ containing gas mixtures was studied in \cite{scott_sources_1982} and \cite{seguin_ultraviolet_1974} with the purpose of improving the output power of TEA~(Transversely Excited Atmospheric) CO$_2$ lasers by photo-ionization of the gas admixtures. The measurements of Seguin~{\it et al}~\cite{seguin_ultraviolet_1974} for several gas mixtures~(e.g. CO$_2$~-~N$_2$~-~He) indicated that the photo-electron density is reduced by increasing the CO$_2$ fractions. 

In CO$_2$ with an admixture of air, the radiation in $98~-~102.5$~nm range that is emitted by nitrogen molecules and ionizes oxygen molecules, can be absorbed by CO$_2$. According to figure~24 in \cite{pancheshnyi_photoionization_2014}, the absorption coefficient for such radiation by CO$_2$ is in range of $0.8~-~2$~$\rm{cm}^{-1}\rm{Torr}^{-1}$~($=60~-~150$~$\rm{mm}^{-1}\rm{bar}^{-1}$), corresponding to an absorption length in the range of $6.1~-~17$~$\mu$m at standard temperature and pressure), hence the effect is quite local when considered on the scale of intrinsic streamer lengths. Furthermore, absorption of radiation in this energy range does not lead to ionization, as the CO$_2$ ionization threshold is about 89~nm~(13.8~eV)~\cite{pancheshnyi_photoionization_2014}. 

In oxygen, radiation originating from the dissociative excitation of oxygen can ionize oxygen molecules and may be important for streamers in pure oxygen if the electron energy distribution shifts such that there is a substantial number of electrons with energy above 20~eV. However, similar to the above, CO$_2$ molecules absorb this radiation after some tens of micrometers without being ionized.

More in general, in CO$_2$ dominated mixtures at STP, the photon absorption length is smaller than about 40~$\mu$m, at least in the photon energy range from 7.6 to 16.9~eV and probably up to higher energies, based on the shape of the absorption curve~\cite{pancheshnyi_photoionization_2014}.

We conclude that in CO$_2$ admixed with air, oxygen or other admixtures, photoionization is not a relevant mechanism for streamer propagation because of the short absorption length of relevant photons in CO$_2$. In this paper, for the simulation of positive streamers in CO$_2$, and CO$_2$ admixed with $1\%$ and $10\%$ oxygen, we incorporate a background density of electrons and positive ions. Such a density could be present for example due to
previous discharges in a repetitively pulsed system~\cite{nijdam_investigation_2014}.

\subsection{Transport and reaction parameters}
\label{sec:transport-reaction}

\begin{figure}
  \begin{center}
    \includegraphics[width=\linewidth]{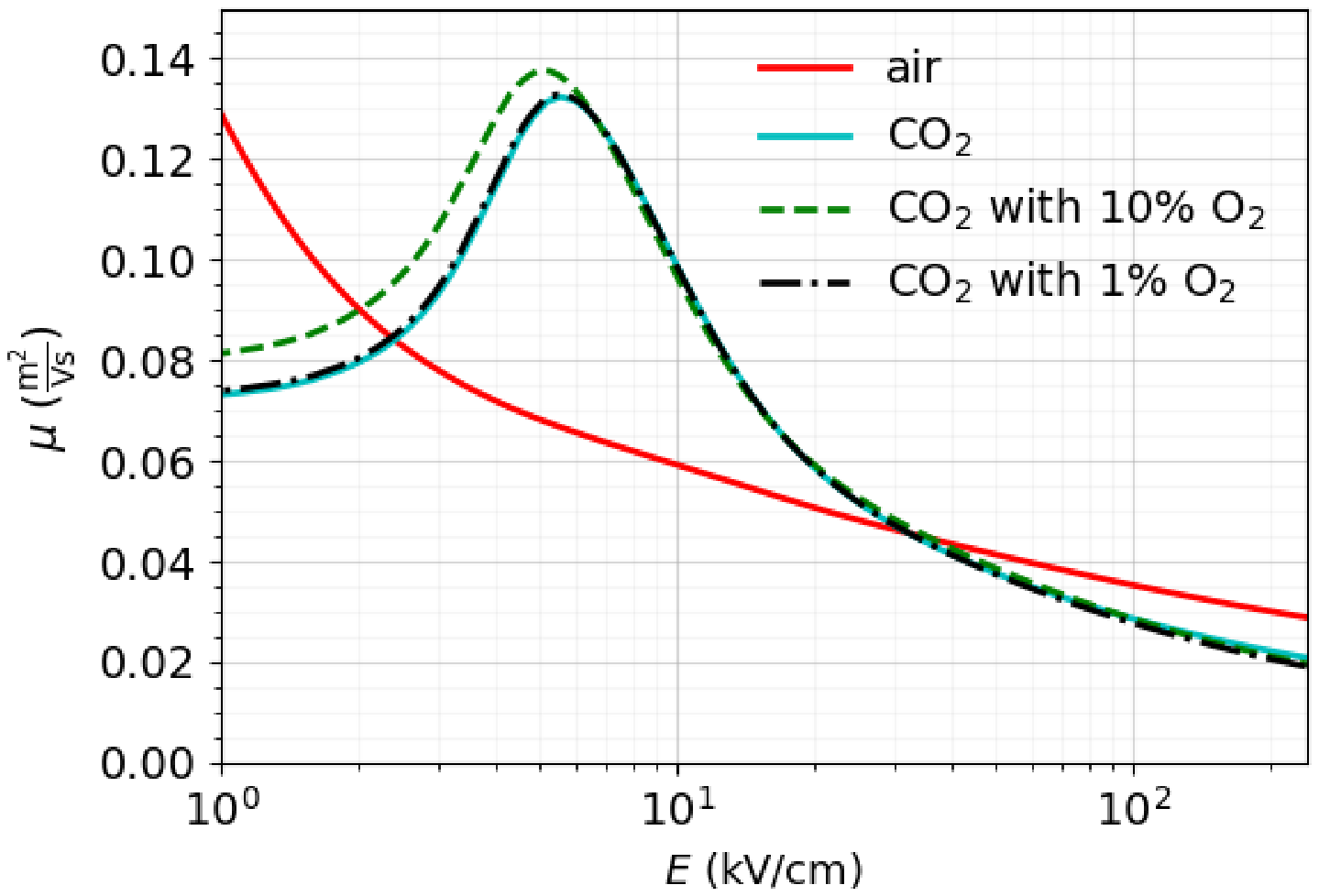}
    \includegraphics[width=\linewidth]{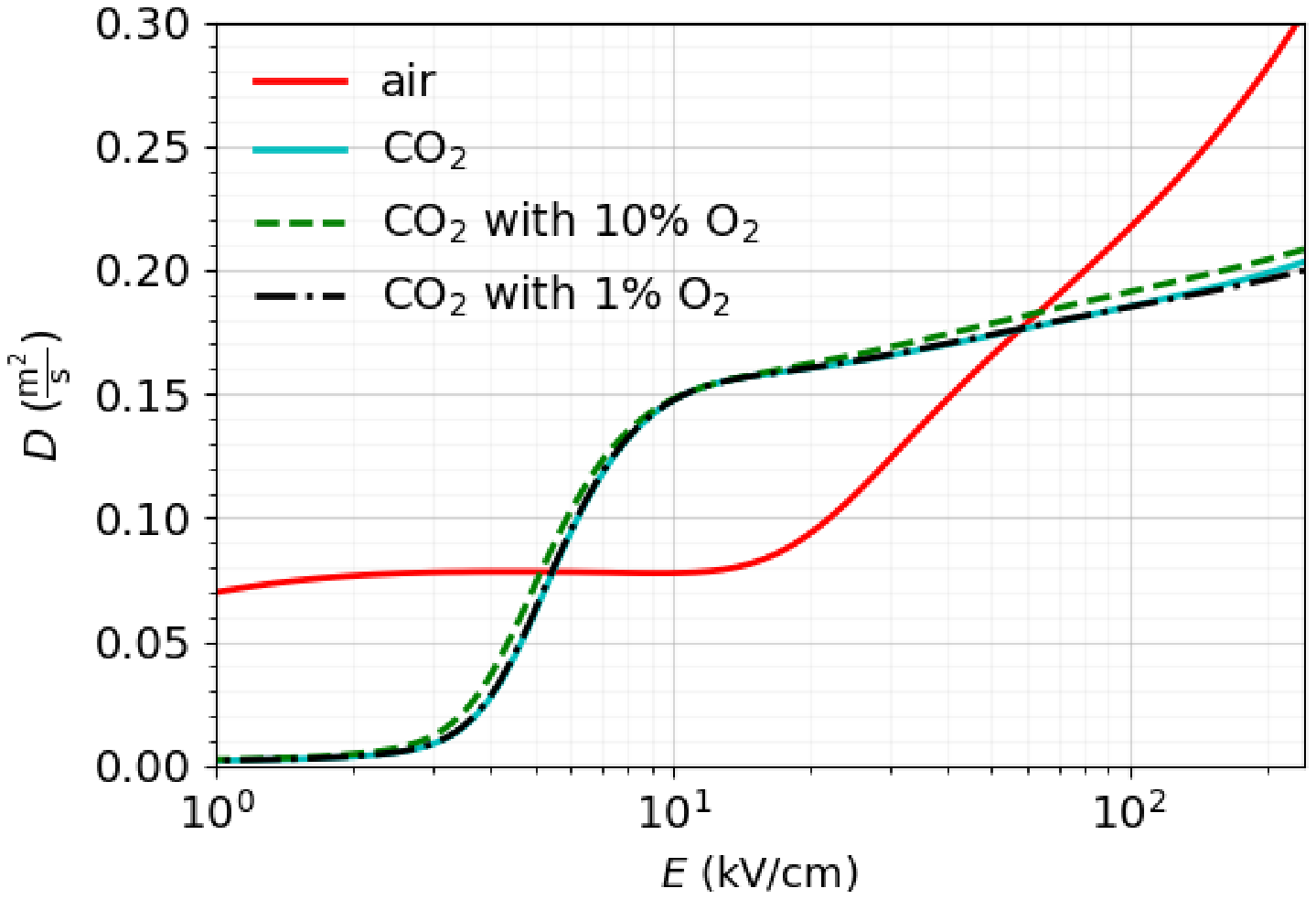}
        \includegraphics[width=\linewidth]{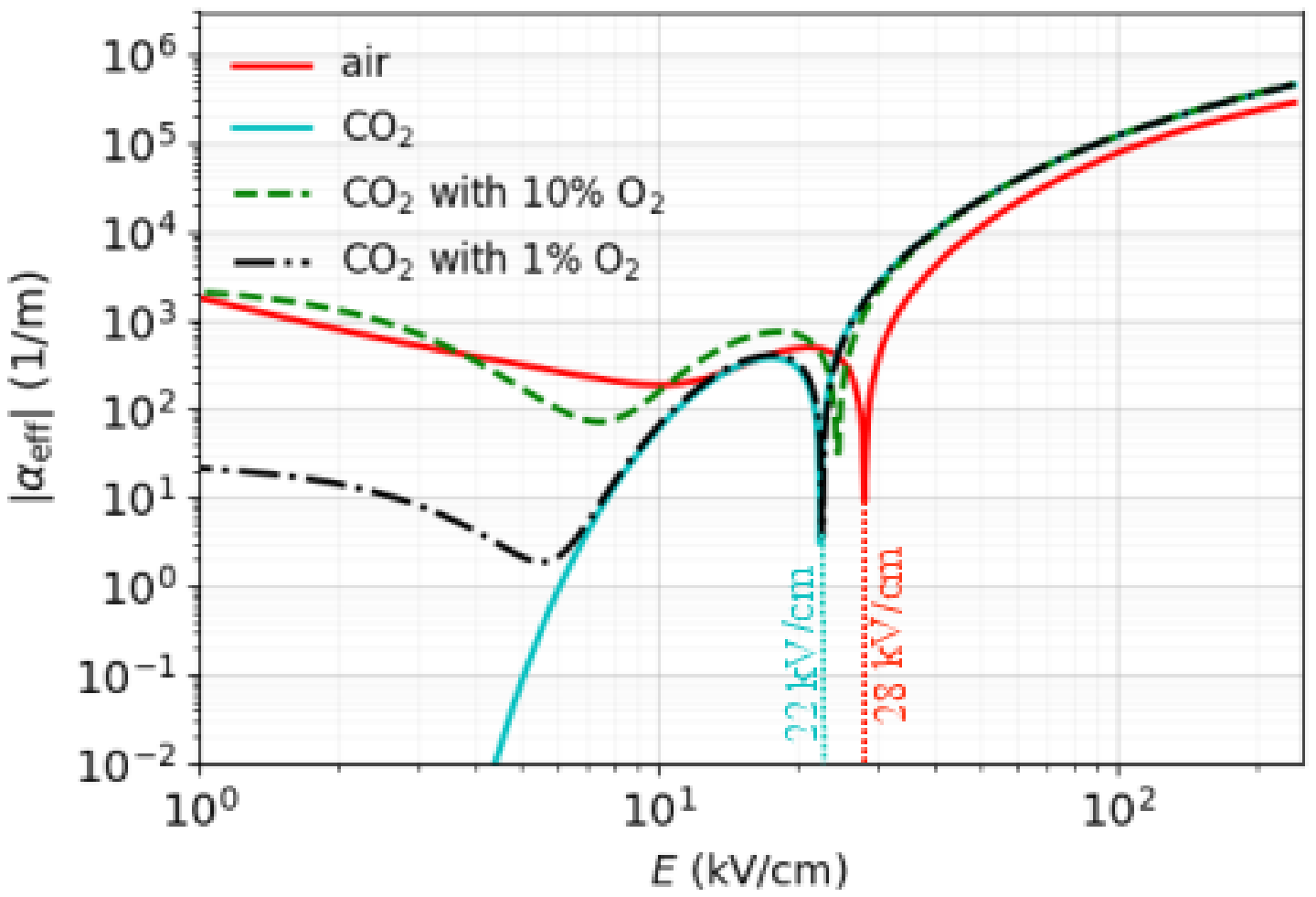}
    \caption{The electron mobility~(top), the diffusion coefficient~(middle), and the effective ionization coefficient~(bottom) at STP condition for air, CO$_2$, and CO$_2$ with 1$\%$ or 10$\%$ of O$_2$. The breakdown field for air is $E^{{\rm air}}_{k}=28$~kV/cm and for CO$_2$ $E_k^{{\rm CO}_2}= 22$~kV/cm.}
    \label{fig:drift-diffusion-coeff-all}
  \end{center}
\end{figure}

Electron-neutral scattering cross sections for CO$_2$ are taken from IST-Lisbon database~\cite{grofulovic2016electron} and for O$_2$ and N$_2$ from Phelps database~\cite{phelps_nodate}, retrieved in May 2019. 
% Jannis removed: Equation~\ref{equ:tau-eff}, gives the relation between $k_q$ and $p_q$.
All transport and tabulated rate constants are calculated with BOLSIG+~\cite{hagelaar_solving_2005}, using the default temporal growth model.

The electron mobility~$\mu_e$, the diffusion coefficient~$D_e$ and the effective ionization coefficient~$|\alpha_{\mathrm{eff}}|$ at standard temperature and pressure are plotted in figure~\ref{fig:drift-diffusion-coeff-all} for all gas mixtures considered in this paper.
Here the effective ionization coefficient is defined as  $\alpha_{\mathrm{eff}}=\alpha-\eta$, where
$\alpha=S_i/(n_e\mu_e|\bf{E}|)$ is the impact ionization coefficient and $\eta=S_{\rm{attach}}/(n_e\mu_e|\bf{E}|)$ the attachment coefficient. 

For CO$_2$ the electron mobility can be seen to be maximal at around 5.5~kV/cm~\cite{Frank2018}) and almost twice as high as in air. In the range of 5.5~kV/cm to 60~kV/cm, electrons in CO$_2$ have a higher diffusion coefficient than in air. Moreover, one can observe that the effective ionization coefficient in CO$_2$ is slightly higher than in air. The breakdown field, defined as the field where $\alpha=\eta$, is around 22~kV/cm for CO$_2$ and 28~kV/cm for air, both at STP. By including $10\%$ or $1\%$ of O$_2$ into CO$_2$, the overall behavior of the parameters does not change, however the values are slightly changed. This shows that the reaction and transport coefficients in the studied gases are dominated by the majority molecule, CO$_2$.

\subsection{Computational domain and initial conditions}
\label{sec:comp-domain}

\begin{figure}
  \centering
  {\includegraphics[width=0.30\textwidth,height=\textheight,keepaspectratio]{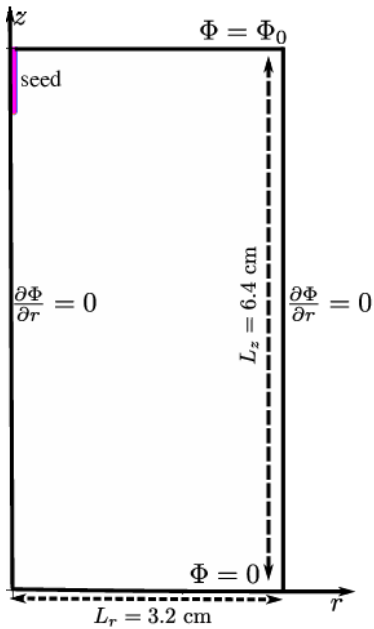}}
  \caption{The axisymmetric domain~ extends over $0 \le r \le L_{r}=3.2$~cm and $0\le z \le L_{z}=6.4$~cm. The position of the
    neutral seed on the axis of symmetry ($r=0$) is indicated; the streamer starts from here. %The seed is about 3.84~mm long with a radius of about 0.02~mm. It has an electron and positive ion density of $5\cdot10^{19}$~$\rm{m}^{-3}$ at its center, which decays at distances above $d = 0.1$~mm with a smoothstep profile: $1-3x^2 + 2x^3$
%, where $x = (d-0.1~{\rm mm})/0.1~{\rm mm}$. 
The boundary at $z=6.4$~cm is on a potential $\Phi=\Phi_0$ and the boundary at $z=0$~cm
is grounded. For the potential, 
Neumann zero boundary conditions are used
at $r=0$~(imposed by symmetry), and at $r=L_{r}$. Neumann zero boundary conditions are also applied for the electron density on all
boundaries. }
  \label{fig:computational-domain}
\end{figure}

The computational domain shown and described in figure~\ref{fig:computational-domain} is used for axisymmetric
simulations. A potential difference of %, where $z_{\rm{max}}=3.2$~cm, and $r_{\rm{max}}=1.6$~cm. 
%The boundary at $z=0$~cm is at a potential 
$\Phi_0$ is applied over a distance of $z=6.4$~cm, creating a homogeneous background electric field. In this paper we employ different values for $\Phi_0$. This leads to different background electric fields, which are indicated explicitly in each section.

%For the potential, Neumann zero boundary conditions are used at $r=0$, and $r=r_{\rm{max}}$. 

%Neumann zero boundary conditions are also applied for the electron density on all boundaries. \hl{Jannis: the boundary conditions are now described both in the figure caption and in the text, maybe refer to the figure in the text?} 

We place a neutral seed of about 7.68~mm long with a radius of about 0.02~mm at the top boundary on the symmetry axis, $(r,z)=(0,6.4~\rm{cm})$. The seed has an electron and positive ion density of $5\cdot10^{19}$~$\rm{m}^{-3}$ at the center. This density decays with a smoothstep profile as $1-3l^2 + 2l^3$, where
$l = \max\left[0, d/(0.1~{\rm mm})-1\right]$ and $d$ is the distance to the line segment defining the seed.

%A homogeneous background density of
%$n_e=n_i=10^9 \, \textnormal{m}^{-3}$ electrons and positive ions is included.
%% Furthermore, an axisymmetric domain with $r_{\mathrm{max}} = z_{\mathrm{max}} = 1.25$~cm is used in which all the other conditions are the same as above.
%For the axisymmetric simulations, we use the same computational domain as in
%\cite{bagheri_comparison_2018}.

In this paper, we use the same refinement criterion as in \cite{teunissen_simulating_2017}. The grid is refined if $\alpha(1.2 \times E) \, \Delta x > 0.8$, where $\alpha(E)$ is
the field-dependent ionization coefficient, $E$ is the electric field strength, and $\Delta x$ is the grid spacing. This gives an AMR grid with a minimum grid spacing of
around $2 \, \mu\textrm{m}$.
%%% Local Variables:
%%% mode: latex
%%% TeX-master: "Paper+Ute"
%%% End:

\section{Results and discussion}
\label{sec:results}

In section \ref{sec:posit-stre-co2}, we first discuss how the absence of
effective photoionization in CO$_2$ can affect positive streamers, due to the lack
of free electrons ahead of them. Afterwards, we compare streamer properties in
air and CO$_2$ when a sufficient number of free electrons is available due to
either background or photoionization. A first overview of results is given in
section~\ref{sec:comparison-propagation}. Several topics are then studied in
more detail. In section \ref{sec:air}, we compare the effect of photoionization
versus background ionization in air. Next, the effect of different background
ionization levels in pure CO$_2$ is studied in section \ref{sec:CO2}, after
which we compare the results in air and CO$_2$ in more detail. Finally, we
present streamer simulations in CO$_2$ with an admixture of 1 or 10\% of oxygen.

\begin{figure*}
  \begin{center}
     \includegraphics[width=\linewidth,height=0.5\textheight,keepaspectratio]{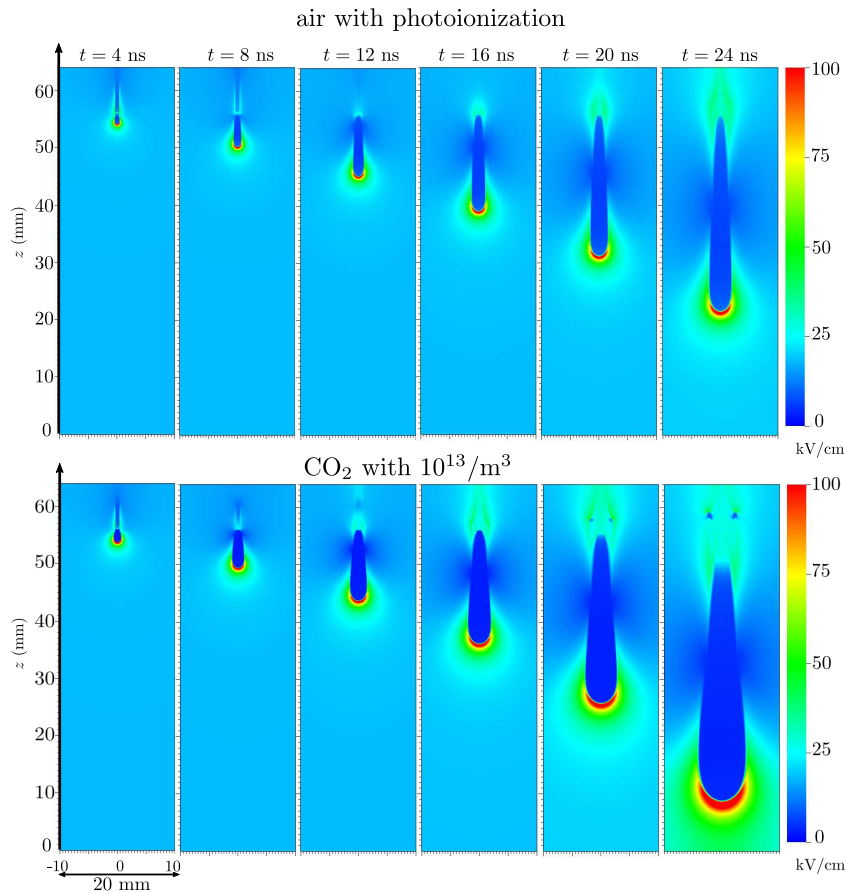}
       \caption{Time evolution of the electric field for the streamer in air (top) and in CO$_2$ (bottom) at standard temperature and pressure in a homogeneous electric field of 18~kV/cm. In the simulation of the air streamer, photoionization is included, whereas in the CO$_2$ streamer, a background electron density of $10^{13}$/m$^3$ is incorporated. The full gap length of 64~mm is shown. The simulation domain extends up to 32~mm in the radial direction, but only 10~mm are shown. The color-coding of the electric field strength is truncated for values above 100~kV/cm.}
    \label{fig:time-evolution}
  \end{center}
\end{figure*}

\begin{figure}
  \begin{center}
     \includegraphics[width=\linewidth]{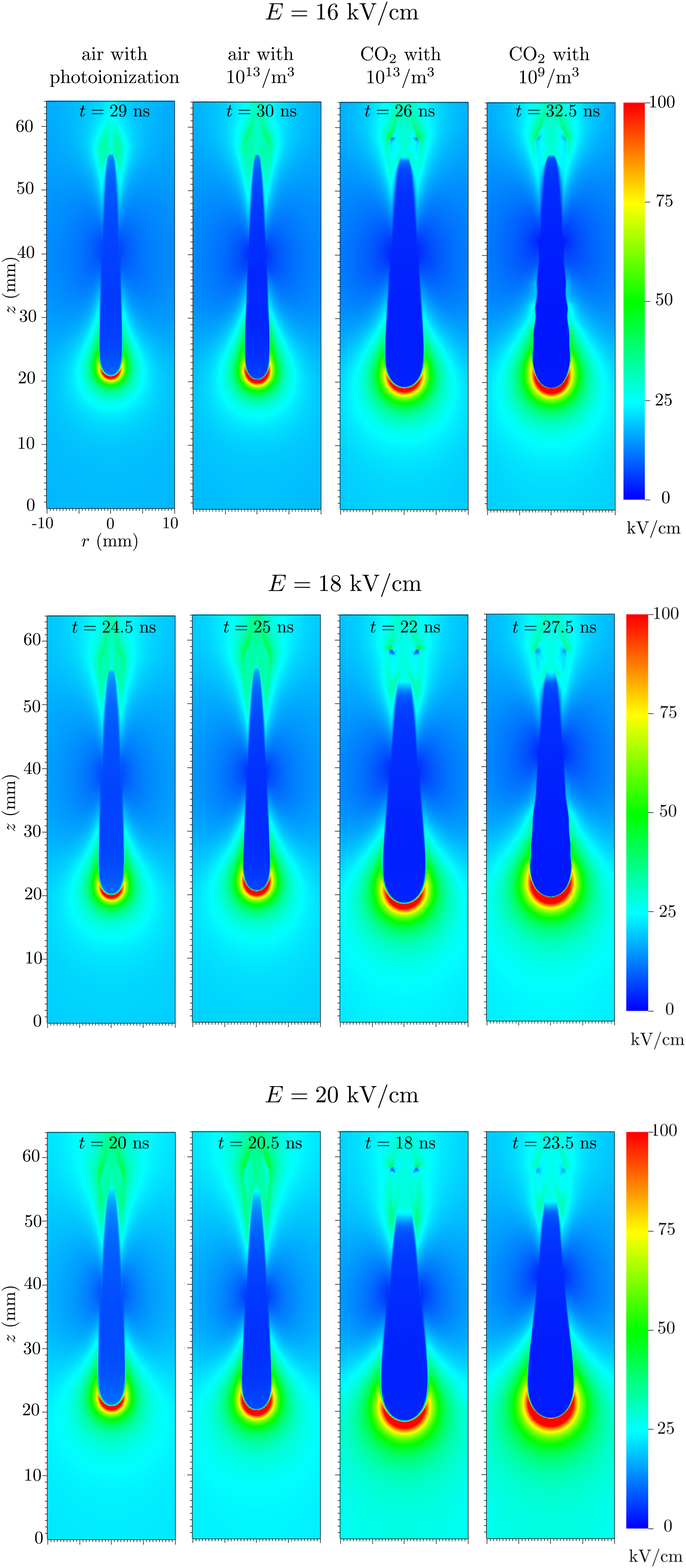}
       \caption{The electric field profile of the streamers in air and CO$_2$ when the streamer head is approximately at the coordinate $z=20$~mm.
       The time when this coordinate is reached, is indicated at the top of each panel. The background electric field is $E=16$~kV/cm~(top), $E=18$~kV/cm~(middle), and $E=20$~kV/cm~(bottom). The inclusion of photoionization or background ionization is indicated above each column. The simulation domain extends up to 32~mm in the radial direction, but only 10~mm are shown.}
    \label{fig:field}
  \end{center}
\end{figure}

\subsection{Positive streamers in CO$_2$ without background ionization}
\label{sec:posit-stre-co2}

An important conclusion from section \ref{sec:photoionization} is that there
seems to be almost no photoionization in pure CO$_2$ or in CO$_2$ with a small
admixture of oxygen or air. This could have a strong effect on positive
streamers in such gases, since their growth depends on the presence of free
electrons ahead of them. Such free electrons can also be provided by background
ionization, \ue{e.g., by electron detachment from negative ions or by external radiation. However, without such electron sources}, background
ionization levels will be low~\cite{pancheshnyi_role_2005}. In such cases, we
expect several observable effects on positive streamers.
\begin{itemize}
  \item The growth of the streamers would be highly irregular, as there would be
  few incoming electron avalanches, leading to a branched structure.
  \item Perhaps, the few incoming avalanches could become tiny negative
  streamers each extending the positive channel.
  \item The resulting discharge would have sharp features, leading to high local
  electric fields and an increased degree of ionization.
\end{itemize}
We are not aware of direct experimental evidence for such effects, probably
because streamers in CO$_2$ emit little visible light. Some of the above effects
have been observed in other gas mixtures with less photoionization than
air~\cite{nijdam_probing_2010,teunissen_3d_2016}, such as N$_2$ with a small
admixture of O$_2$. Note that a distinguishing property of CO$_2$ is that it is both electronegative (unlike
e.g. N$_2$ or Ar) and that it strongly absorbs the photons responsible for
photoionization in air, ruling out photoionization due to air impurities.

Simulating positive streamers under the conditions outlined above is highly
challenging, and outside the scope of the present paper. In the rest of the
paper we therefore compare streamer properties in different gases with a
sufficient number of free electrons available, \ue{and we test how sensitively our results depend on the assumed electron density}.

\subsection{A first look at streamers in air and CO$_2$}
\label{sec:comparison-propagation}

In this section, we have a first look at streamer properties when a sufficient
number of free electrons is available ahead of them. In CO$_2$ we provide such
free electrons by adding a certain level of background ionization.

Figure~\ref{fig:time-evolution} shows the dynamics of the streamer evolution in air~(top) and in CO$_2$~(bottom) in a homogeneous electric field of 18~kV/cm. For the CO$_2$ streamer, a background density of free electrons and positive charges of 10$^{13}$/m$^3$ is incorporated, whereas for the air streamer photoionization is included. Initially the electric field is enhanced at the location of the seed, and within a couple of nanoseconds a positive streamer propagates downwards. The air streamer bridges the gap after 30~ns and the CO$_2$ streamer after 25~ns. In this paper, we focus on the streamer propagation far from the electrodes, and we stop before the streamer has reached the opposite electrode. 

Figure~\ref{fig:field} shows the electric field profile of the streamers in air and CO$_2$ when they have reached a fixed length. In the top row of panels a homogeneous electric field of 16~kV/cm is applied, in the middle row the field is 18~kV/cm and in the bottom row, it is 20~kV/cm. The results of air streamers with photoionization are included in the first column. In the other three columns, results in air and CO$_2$ with background ionization are shown, as indicated. To see the differences between the streamers more clearly, the electric field  profiles are shown at the same streamer length, but at different times; these times are indicated in the top of each panel.

In what follows, we provide a detailed analysis of the properties of these streamers in air and in CO$_2$.

\subsection{Streamers in air: photoionization versus background ionization}
\label{sec:air}
In this section, we investigate the influence of photoionization versus background ionization and of the applied electric field on streamer properties in air. We performed two sets of simulations, where all the conditions are identical except that in one case we included photoionization in continuum approximation using Bourdon's three term parameters as described in section~\ref{sec:photoi-model}, and in the other case we incorporated a background ionization of $10^{13}$/m$^3$, but no photoionization.

\begin{figure}
  \begin{center}
   \includegraphics[width=\linewidth]{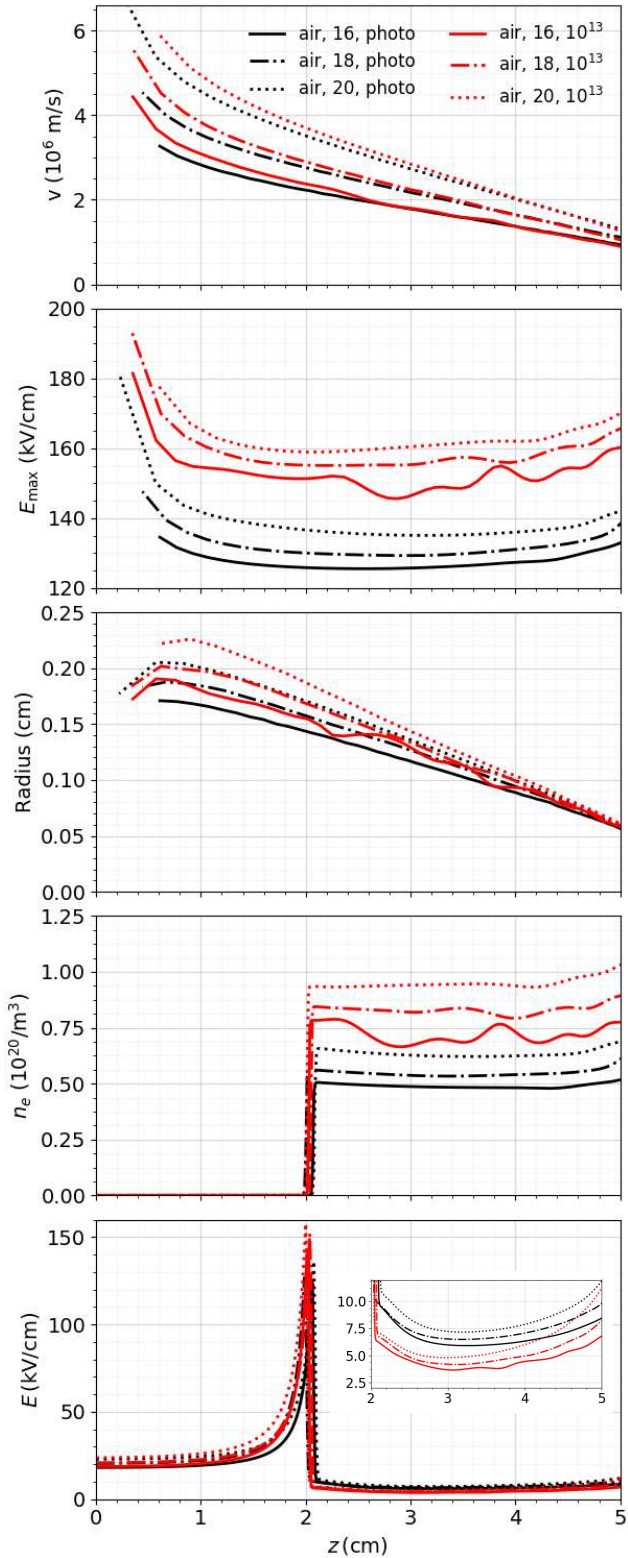}
    \caption{From top to bottom: streamer velocity, maximal electric field, and radius of the streamer head as a function of streamer position $z$; and electron density and electric field on the $z$ axis at the moment when the streamer head is approximately at $z=2$~cm.
    The simulations are done here in dry air. The different lines indicate results with photoionization or with a background electron density of $10^{13}$/m$^3$ in applied electric fields of 16, 18 and 20~kV/cm.}
    \label{fig:E-L-v-1}
  \end{center}
\end{figure}

\jt{The upper three panels in Figure~\ref{fig:E-L-v-1} show the streamer velocity, maximal electric field, and radius versus streamer position for cases with photoionization or background ionization in three different applied electric fields.}
\jt{Here, the streamer position is defined as the $z$-coordinate where the electric field is maximal, and the streamer radius is defined as the radius where the radial component of the electric field is maximal. Note that these definitions are used throughout this paper.}
\ue{The initial transients during streamer formation for $z>5$~cm are not shown in the figure, and the streamer propagates towards the lower electrode at $z=0$. Boundary effects from approaching the electrode can be seen for $z\le 1$~cm.} 
\jt{The lower two panels in Figure~\ref{fig:E-L-v-1} show on-axis electron density and electric field profiles along the $z$-axis when the streamer head is approximately at $z = 2 \, \textrm{cm}$. These two panels correspond to the left two columns of figure~\ref{fig:field}, where the times are indicated.} 
The following observations can be made in the figure: 

\paragraph{Velocity.} As the streamers propagate, their velocities increase from $1\times10^6$~m/s to about $6\times10^6$~m/s. The streamer velocities coincide well until \ue{$z\approx$ 3~cm. Then streamers with background ionization become faster in each external electric field, up to about $6\%$ when approaching the opposite electrode.} Moreover, the higher the electric field the faster the streamer travels.

\paragraph{Maximal electric field.} \ue{During the initial streamer formation from the ionization seed, the maximal electric field briefly reaches about 200~kV/cm; this occurs for $z>5$~cm and is not shown in the figure.  Then,} during the extended streamer propagation phase, the maximal electric field at the streamer head is about $145-160$~kV/cm in cases with background ionization, and about $125-135$~kV/cm in cases with photoionization. Hence, with background ionization the field is by about $20\%$ larger. When the applied electric field is larger, the maximal field at the streamer head is larger as well for each streamer length. \jt{Oscillations are visible in the electric field (and other quantities) for the case with the lowest field and background ionization; these oscillations are discussed below in section \ref{sec:oscillations}.}

\paragraph{Radius.} The streamer radius increases from about 0.5~mm to about 2.2~mm as it propagates between the two electrodes. It is about $10\%$ larger in streamers with background ionization. Furthermore, for a larger applied electric field the radius is larger for each streamer length.

\paragraph{Electron density in the streamer interior.}
The streamers with background ionization have a larger electron density in the
streamer interior than those with photoionization. They also have a higher maximal
electric field at the streamer head. That a higher electric field at the tip
creates a higher interior electron density, is established for negative
streamers~\cite{li_deviations_2007}, but will require further investigations for
positive streamers. \jt{There appear to be several competing effects here.
  Photoionization is strongest on-axis, which `focuses' the growth of a positive
  streamer and therefore could explain the smaller radius with photoionization.
  Usually, a streamer with a smaller radius will have a stronger electric field
  at its tip. However, here the wider streamers with background ionization have
  stronger electric field enhancement. This is due to their higher electron
  density and thus also higher conductivity, for which we currently do not have
  a simple explanation.}

\paragraph{Electric field in the streamer interior.}
The interior electric field on the axis is in the range of $4-5$~kV/cm in simulations with background ionization and about 2~kV/cm larger in simulations with photoionization.
Interestingly, the applied electric field has a very minor effect on the interior electric field. Apparently, the higher electron density in the interior supports larger screening currents that compensate for the higher fields at the tip. This is a topic of future investigations. 
The electric field of 5~kV/cm is sometimes attributed to the so called `{\it stability field}' for positive  streamers in air. 

\paragraph{We conclude} that the replacement of photoionization by a background electron density of $10^{13}/$m$^3$ in air provides a qualitative description of streamer properties.
Both mechanisms create a sufficiently similar electron density profile in the active high field zone ahead of the streamer head within the parameter range explored in this paper. 

\subsubsection{Oscillations}
\label{sec:oscillations}

In figure \ref{fig:E-L-v-1} oscillations are visible in all quantities for the 16~kV/cm case with background ionization.
% These oscillations are visible in its maximal electric field, radius, and internal electron density, but they are not visible in color-coded plots of electric field~(see figure~\ref{fig:field}).
Such oscillations were also observed in~\cite{bagheri_comparison_2018} when a background ionization level of 10$^9$/m$^3$ was used to compare axisymmetric streamer codes. They were attributed to numerical effects as they could be removed by using a very fine grid and corresponding small time step. However, in our present simulations these oscillations did not disappear by reducing the grid size. After an extensive search, we did observe a sensitivity on the number of points in the tabulated input data. From this data, rate coefficients are determined by linear interpolation, which leads to small interpolation errors since processes like electron impact ionization are not linear in $E/N$. To reduce such interpolation errors, high-resolution input data with 200 entries was used for all simulations presented in this paper, but as shown in figure \ref{fig:E-L-v-1}, some oscillations were nevertheless present.

The fact that small numerical or interpolation errors can cause oscillations in positive streamers indicates that these streamers are also `{\it physically}' unstable to some degree.
Our results show that this instability is enhanced when the applied field is reduced. A reduction in the background ionization level to $10^{9}$/m$^3$ (like in~\cite{bagheri_comparison_2018}) also led to significantly increased oscillations and branching. Since the axisymmetric fluid model used here is not suitable for the study of stochastic fluctuations or branching, we leave a further investigation of these effects to future work.

\subsection{Streamers in CO$_2$: different levels of background ionization}
\label{sec:CO2}
In this section, we characterize streamers in CO$_2$. We studied the effect of different levels of background ionization on streamer properties to explore the sensitivity of the results to this parameter. In one set of simulations we included background ionization of $10^9$/m$^3$, and in another set background ionization of $10^{13}$/m$^3$. Streamers in CO$_2$ with background ionization of $10^9$/m$^3$ are more stable than in air; oscillations occur but the streamers do not branch. 

\begin{figure}
  \begin{center}
     \includegraphics[width=\linewidth]{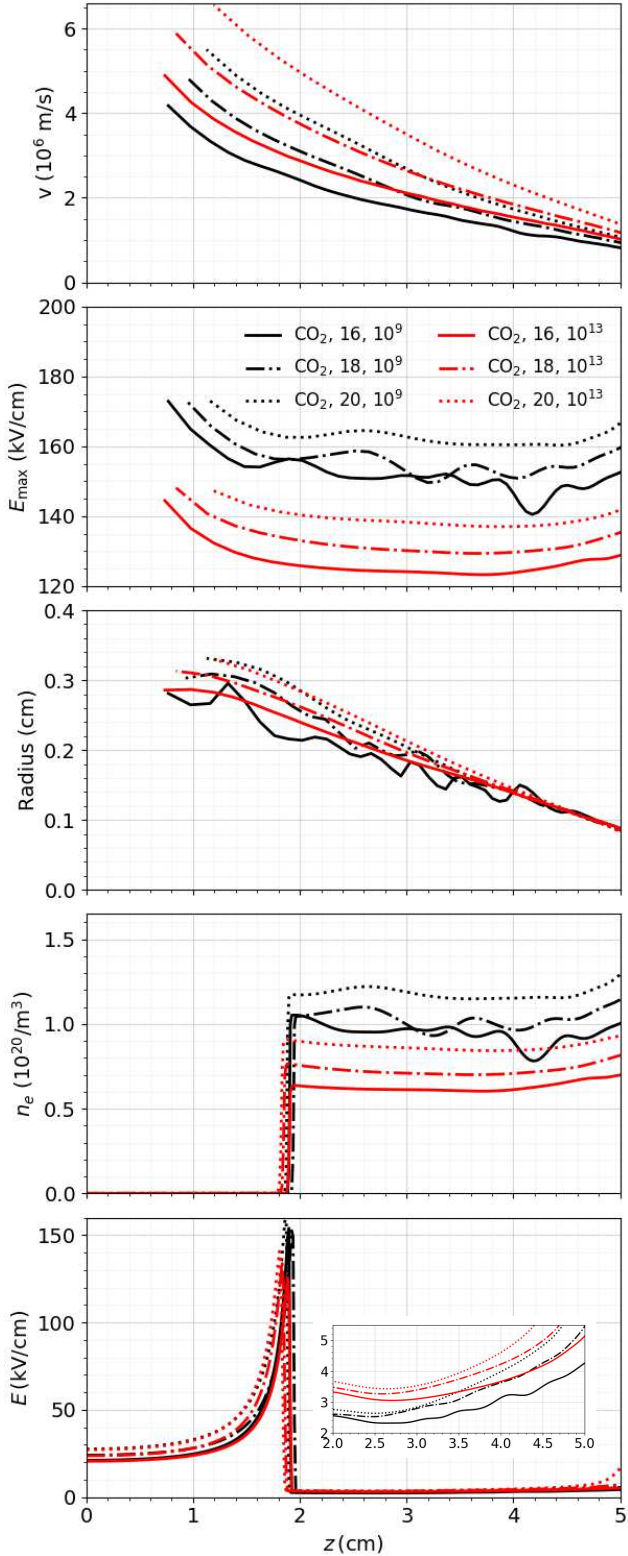}
    \caption{The same plots as in figure~\ref{fig:E-L-v-1}, now for CO$_2$. The different lines indicate results with background ionization of $10^9$/m$^3$ and 10$^{13}$/m$^3$ in applied electric fields of $16$, $18$, and $20$~kV/cm.}
    \label{fig:E-L-v-2}
  \end{center}
\end{figure}

In figure~\ref{fig:E-L-v-2} the same functions are plotted as in figure~\ref{fig:E-L-v-1}, but now for streamers in CO$_2$ with a background ionization of $10^9$/m$^3$ or $10^{13}$/m$^3$. The same initial conditions are used and the same three electric fields are applied.
The profiles of electron density and electric field on axis in the lower two panels of figure~\ref{fig:E-L-v-2} show the same situation as the right two columns in figure~\ref{fig:field}. The following observations can be made: 

\paragraph{Velocity.} The streamer velocity increases in time from about $(1-1.5)\times10^6$~m/s to $(4-4.5)\times10^6$~m/s depending on the level of background ionization and on the applied electric field. By increasing the applied electric field from 16 to 20~kV/cm, the streamer velocity increases by up to 50~\% for each streamer length. When the background electron density is increased by 4 orders of magnitude, the streamer velocity for given streamer length varies by less than 10~\%, i.e., it is very insensitive to such a large change.

\paragraph{Maximal electric field.} During the streamer propagation phase, the maximal electric field at the streamer head is about $125-140$~kV/cm in cases with background ionization of $10^{13}$/m$^3$ and it is about $150-165$ in the cases with background ionization of $10^{9}$/m$^3$, i.e., it increases by about 25~\% when the background electron density is reduced, but only by about 10~\% when the applied electric field is increased.

Note that the maximal fields for streamers in applied fields of 16 and 18~kV/cm, and with background ionization of $10^{9}$/m$^3$ strongly oscillate. Such oscillations are discussed in section~\ref{sec:oscillations}. 

\paragraph{Radius.} The streamer radius increases from about 0.8~mm to about 3.5~mm in time. It is somewhat higher for the higher background ionization and the higher applied electric fields.

\paragraph{Electron density in the streamer interior.}
As already said above, the electron density behind a negative streamer ionization front is determined by the maximal electric field at the streamer head~\cite{li_deviations_2007}. A similar relation can be seen here for the positive streamers: the internal electron density depends more strongly on the shown levels of background electron density and more weakly on the applied electric field. This is the same functional dependence as for the maximal electric field discussed above.

\paragraph{Electric field in the streamer interior.}
The electric field on the axis is in the range of $2-3$~kV/cm for $10^{9}$/m$^3$, and of $3-4$~kV/cm for $10^{13}$/m$^3$. Again it only weakly depends on the applied electric field, but more strongly on the background electron density level.

\paragraph{Our main conclusion} is that the streamer properties change only by up to 50~\% and frequently much less, when the background electron density provided at the start of the simulation is changed from $10^9$ to $10^{13}$/m$^3$, i.e., by 4 orders of magnitude. As a proper estimate of such a density is challenging, it is useful to note that the results on streamer propagation are rather insensitive to this parameter.

\subsection{Air streamers versus CO$_2$ streamers}
\label{sec:airvsCO2}
In this section, we compare streamers in air with those in CO$_2$.
A first view is already given in figure~\ref{fig:field}, where the CO$_2$ streamers are wider than air streamers for all values of the electric field. Furthermore, for the same background electron density and applied electric field, the CO$_2$ streamers are faster and they have a lower maximal electric field at their head.
However, the fact that the CO$_2$ streamer are faster and wider could be due to the fact that the breakdown field $E_k$ in air is 28~kV/cm and in CO$_2$ only 22~kV/cm. The fixed electric fields of 16 to 20~V/cm of the previous simulations are therefore closer to the breakdown field of CO$_2$.

\begin{figure}
  \begin{center}
  \includegraphics[width=\linewidth]{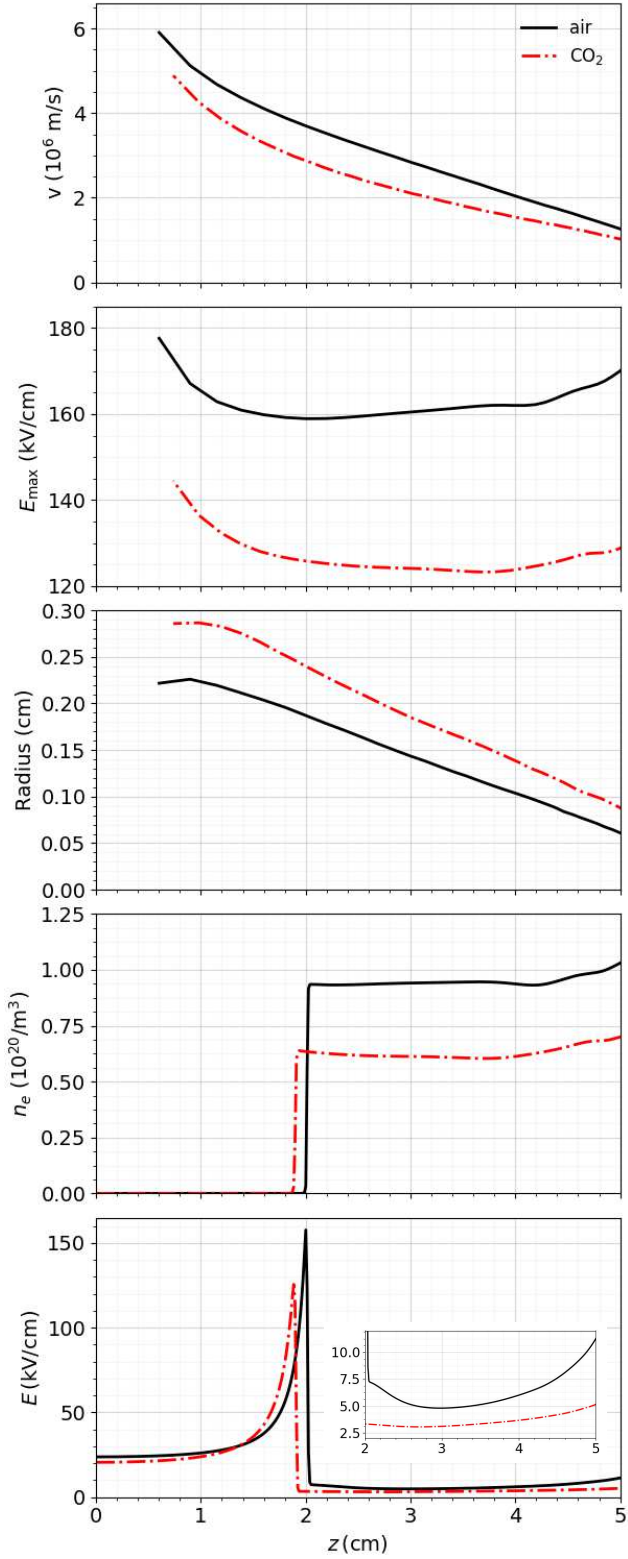}
    \caption{The same plots as in figure~\ref{fig:E-L-v-1}, now for air and CO$_2$ with background ionization of 10$^{13}$/m$^3$ in an applied electric field of $E=0.73\,E_k$. }
    \label{fig:E-L-v-4}
  \end{center}
\end{figure}

Therefore, we here present simulations in air and in CO$_2$ at the same fraction $E=0.73\,E_{k}$ of their respective breakdown fields 
(hence for $E=16$~kV/cm for CO$_2$ and $E=20$~kV/cm for air),
and with the same background electron density of 10$^{13}$/m$^3$.
Figure~\ref{fig:E-L-v-4} shows the same plots as the previous two figures for these two gases.
According to Figure~\ref{fig:field}, the air streamer then has propagated for 20.5~ns, and the streamer in CO$_2$ for 26~ns.

Hence the air streamer now propagates about 30~\% faster than the CO$_2$ streamer. The maximal electric field at the tip of the air streamer is about 35~kV/cm larger. And the streamer radius in air is about 15~\% smaller.
The interior electron density is larger in air than in CO$_2$. The interior electric field of air streamer reaches to a minimum value of about 5~kV/cm, whereas the interior electric field of the CO$_2$ streamer becomes as low as about $2.5$~kV/cm. 

\subsection{Streamers in CO$_2$ with an oxygen admixture of 1$\%$ or $10\%$}
\label{sec:CO2-O2}
\begin{figure}
  \begin{center}
     \includegraphics[width=\linewidth]{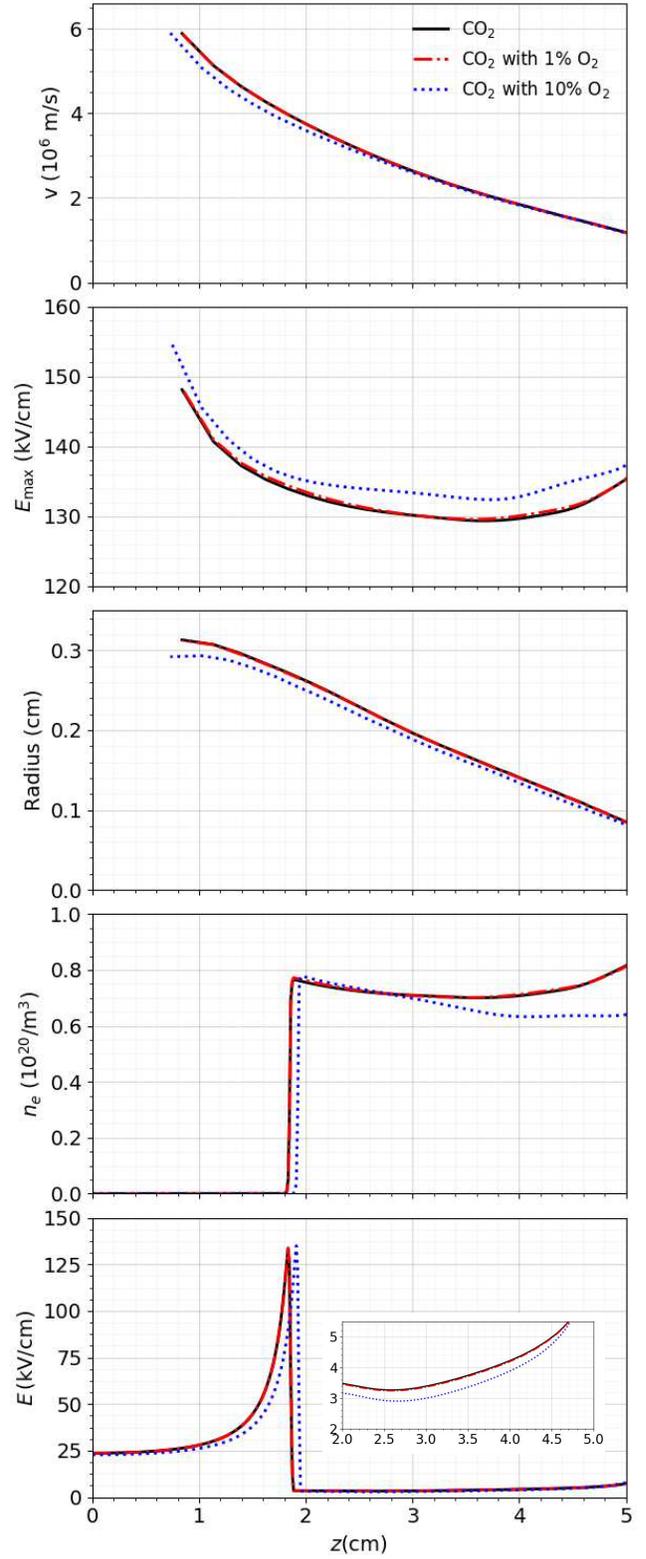}
    \caption{The same plots as in figure~\ref{fig:E-L-v-1}, now for pure CO$_2$ and for CO$_2$ with $1\%$ or $10\%$ of oxygen admixture. The applied electric field is E=18~kV/cm, and background ionization is $10^{13}$/m$^3$.}
    \label{fig:E-L-v-5}
  \end{center}
\end{figure}

As we mentioned in the introduction, in a circuit breaker, an admixture of O$_2$ is used to suppress soot formation in a CO$_2$ discharge. In this section, we investigate the effect of oxygen admixture of 10$\%$ and $1\%$ on streamer properties in CO$_2$. We performed simulations using a background ionization of $10^{13}$/m$^{3}$ and an applied electric field of 18~kV/cm. Figure~\ref{fig:E-L-v-5} shows similar quantities as previous sections: the streamer velocity, maximal electric field, radius, and on-axis electron density and electric field profiles when the streamer is at about $z=2$~cm. Streamer properties in CO$_2$ essentially do not change with an oxygen admixture of $1\%$. By increasing the oxygen admixture to $10\%$ some small deviations start to appear. Most notable is the decay of the electron density on the streamer axis behind the ionization front in the case of the 10\% O$_2$ admixture.
This is due to the higher electron attachment rate (shown in Figure~\ref{fig:drift-diffusion-coeff-all}) in the streamer interior where the field is below 4~kV/cm. 

%\begin{figure}
%  \begin{center}
%     \includegraphics[width=\linewidth]{Screenshot.png}
%    \includegraphics[width=\linewidth]{Screen2.png}
%     \caption{Just for info: O$_2^{-}$ density. Color-codes are the same.}
%     \label{fig:extra-info}
%    \end{center}
%  \end{figure}

%%% Local Variables:
%%% mode: latex
%%% TeX-master: "Paper+Ute"
%%% End:

\section{Conclusion and outlook}
\label{sec:conclusions}

We have presented simulations of the evolution of positive streamers in CO$_2$ and in air, with an emphasis on velocity, radius and maximal electric field at the streamer head, and on the generated electron density profile and the electric field inside the streamer channel.

\subsection{Lack of photoionization in CO$_2$ containing gases}

%MY FOLLOWING PARAGRAPH IS LONG AND EXPLANATORY, AND WE COULD SWITCH TO A SHORTER VERSION GIVEN FURTHER BELOW, AND USE THIS ONE IN SECTION 2. WHAT DO YOU THINK?
% \hl{Jannis: I would split off the first part, which does not fit with the subsection title. Done by Behnaz and Ute!}

A major challenge for understanding positive streamers in CO$_2$ is that one needs a source of free electrons ahead of the ionization front for the streamer to propagate. In air, it is well established that photoionization provides such a source: for the typical electron energy distribution in a streamer ionization front, a wave length band of photons is generated that has sufficient energy to ionize an oxygen molecule, and that can propagate a distance of the order of a millimeter through air at standard temperature and pressure without being absorbed. Hence, photoionization in air provides a source of free electrons extending up to millimeters ahead of the ionization front. 
However, as we have reviewed in section 2, no such non-local electron source is known in CO$_2$. Rather, the CO$_2$ molecule absorbs photons in the relevant energy range after a few tens of micrometers. An admixture of oxygen or air does not help either as the photons relevant for the photoionization in nitrogen-oxygen mixtures are strongly absorbed by CO$_2$ as well.

A possible conclusion from this lack of non-local photoionization is that positive streamers in pure CO$_2$ or in CO$_2$ with admixtures of nitrogen and oxygen do not propagate at all, if there is no alternative source of free electrons ahead of the front. The consequences of such a lack of free electrons ahead of the streamer are described in section~\ref{sec:posit-stre-co2}. As in air, such a source could be some background ionization due to previous discharges, or due to radioactive admixtures or other sources of external radiation.

\subsection{Photoionization versus background ionization}

Rather than searching for such specific sources, we have investigated the sensitivity of streamer simulations to photoionization or background ionization. Surprisingly,  when photoionization in air is replaced by a background ionization of $10^{13}$/m$^3$ of free electrons and positive charges, the observed streamer parameters vary by no more than 20~\% within the parameter range of our simulations. 

Similarly, when we assume a density of $10^9$ or $10^{13}$/m$^3$ of free electrons
and positive charges in CO$_2$, the streamer properties (velocity, radius, maximal field, interior field and interior electron density) change by no more than 30\%. From such a small change on a background electron density difference of 4 orders of magnitude, we conclude that the streamer properties during the propagation phase are rather insensitive to this parameter, hence we do not need to know it with high precision within the parameter range of our simulations. However, we expect streamer inception and branching to depend strongly on this density. %(It should be noted though that the background electron density could be relevant for streamer branching.) 
% \hl{Jannis: this formulation should be stronger IMO. B+U: Now better?}

\subsection{The effect of transport and reaction parameters}

% ONE CAN NOT SIMPLY ARGUE THAT BECAUSE THE ELECTRON MOBILITY INSIDE A CO$_2$ STREAMER IS HIGHER, THE INTERIOR FIELD IS LOWER. HOWEVER, THE VALERIA RELATION ALLOWS TO MAKE THE THOUGHT QUANTITATIVE, AND IT COMPARES VERY WELL WITH THE SIMULATION RESULTS. BUT I SUGGEST TO POSTPONE THIS TO A FUTURE PAPER.

The internal streamer dynamics is characterized not only by this free electron source, but also by the electron mobility $\mu$ and by the effective ionization coefficient $\alpha_{\rm eff}$. The breakdown field is the field where $\alpha_{\rm eff}$ vanishes. However, when the background field is chosen as the same fraction of the breakdown field both in air and in CO$_2$ and when a background electron density of $10^{13}$/m$^3$ and no photoionization is used in both gases, the discharges are still not equal. Rather the air streamers have a larger velocity, a larger maximal electric field at the head, a larger interior electron density and electric field and a smaller radius. This is caused by the different functional dependence of $\mu$ and $\alpha_{\rm eff}$ on the electric field, as shown in Fig.~\ref{fig:drift-diffusion-coeff-all}. An important difference is that the electron mobility is substantially larger in the interior of a CO$_2$ streamer, but somewhat lower in the active high field zone ahead of the streamer.
Furthermore, the electron attachment rate in the interior of a CO$_2$ streamer is substantially lower than in air for fields below 10~kV/cm.

\jt{The non-linear evolution of streamer discharges makes it difficult to directly relate the observed differences to these transport coefficients. For example, a smaller radius and lower interior mobility reduce the conductivity of streamers in air, but the higher interior electron density in air has the opposite effect. The effect of the lower attachment rate in CO$_2$ is more clear: streamers in CO$_2$ will retain their conductivity for longer times/distances. This could partially explain why they obtain a larger radius in our simulations.}

% \hl{Jannis: explanation could be elaborated/clarified a bit here. \\ Ute: I just suggested ABOVE not to elaborate. I don't get the argument really consistent without the Valeria relation.}

\subsection{Quantitative results}

\begin{itemize}
    \item Replacing photoionization by a background electron density in simulations of air streamers \jt{does not drastically change streamer properties}, at least within our sets of parameters.
    
    \item Streamers in air propagate faster than in CO$_2$ in a background electric field of 0.73 times the breakdown field of the respective gas. However, in a background electric field of 18~kV/cm streamers \jt{are faster in CO$_2$ than in air}.
    
    \item The interior electric field in CO$_2$ streamers is about $2-4$~kV/cm, whereas in air it is about $4-7$~kV/cm. The applied electric field has a very minor effect on the interior field. At least in air streamers the inclusion of background ionization instead of photoionization reduces the interior electric field.  
    
    \item The streamer properties in CO$_2$ are essentially unchanged when $1\%$ or $10\%$ of oxygen is admixed.
    
\end{itemize}

\subsection{Outlook}

We list here a number of questions left for future studies:
\begin{itemize}
    \item Did we miss some possible source of free electrons ahead of a positive streamer in CO$_2$? Or can we find experimental observations where such a streamer really does not propagate?
    \item Can we estimate the free electron density in repetitive discharges in CO$_2$ for use in simulations?
    \item Can we derive some more quantitative understanding of the relation between the transport and reaction parameters $\mu(E)$ and $\alpha_{\rm eff}(E)$ and the streamer properties?
    \item We tested background electric fields of 16 to 20~kV/cm where the streamers are expanding and accelerating. Will the same conclusions as above hold in lower electric fields or in longer gaps?
\end{itemize}

%%% Local Variables:
%%% mode: latex
%%% TeX-master: "Paper+Ute"
%%% End:

\section*{Acknowledgments}
We acknowledge numerous fruitful discussion with Martin Seeger and his colleagues at ABB Corp. Res., Baden, Switzerland, as well as with our experimental collaborators at Eindhoven Univ. Techn.
B.B. acknowledges funding through the Dutch STW-project 15052 ``Let CO$_2$ spark!".
\section*{Supplementary material}
The simulation code used in this paper is available at https://gitlab.com/MD-CWI-NL/afivo-streamer. Moreover, the tabulated transport and rate constants, and the input files for generating the results together with the
output files will be provided online, when the paper is accepted. 
%\clearpage
\section*{References}
\bibliography{jannis_bibtex_new,CO2-photoionization}
\end{document}